\documentclass[ twocolumn,
aps,nofootinbib,showpacs,showkeys,preprint
tightenlines
] {revtex4}

\usepackage{epsf,epsfig,subfigure,axodraw,graphicx,amsmath,amssymb}
\usepackage{color}
\newcommand{\dis}[1]{\begin{equation}\begin{split}#1\end{split}\end{equation}}

\newcommand{\ie}{{\it i.e.}\ }

\begin{document}


\title{\Large\bf Quark and lepton mixing angles with a dodeca-symmetry}

\author{Jihn E Kim and
 Min-Seok Seo}
\affiliation{ Department of Physics and Astronomy and Center for Theoretical Physics, Seoul National University, Seoul 151-747, Korea
 }
\begin{abstract}
 The discrete symmetry $D_{12}$ at the electroweak scale is used to fix the quark and lepton mixing angles. At the leading order, the Cabbibo angle $\theta_C$ is  15$^{\rm o}$, and the PMNS matrix is of a bi-dodeca-mixing form giving the Solar-neutrino angle $\theta_{\rm sol}=30^{\rm o}$. Thus, there results the relation $\theta_{\rm sol}+\theta_C\simeq 45^{\rm o}$. Out of discrete vacua, a certain vacuum is chosen for this assignment to be consistent with the dodeca-symmetry. A shift of $\theta_C$ from 15$^{\rm o}$ to 13.14$^{\rm o}$ might arise from a small breaking of the dodeca-symmetry. The spontaneous breaking leading to the required electroweak vacuum is made possible by realizing the electroweak dodeca-symmetry explicitly at a high energy scale. At the vacuum we chose Arg.Det.$M_q$ is nonzero, and hence a solution of the strong CP problem invites a very light axion at a high energy scale. We also comment how the next level corrections can fit the mixing angles to the observed values.
An example realizing this idea needs a symmetry
$SU(3)_c\times SU(2)_L\times U(1)_Y\times D_{12}\times U(1)_\Gamma\times Z_3\times Z_2$.
\end{abstract}


\maketitle

\section{Introduction}\label{sec:Introduction}
In the standard model (SM), the Yukawa couplings and the Higgs potential are not completely fixed yet. Nevertheless, it is a phenomenological virtue that these are general enough to allow the quark and lepton masses and their mixing angles at the observed values \cite{PData08}. Here, the unitary matrices diagonalizing the quark and lepton masses introduce the Cabibbo-Kobayashi-Maskawa (CKM) matrix $V_{\rm CKM}$ \cite{CKM73} and the Pontecorvo-Maki-Nakagawa-Sakata (PMNS) matrix $V_{\rm PMNS}$ \cite{PMNS,NeutrinoReview}. When one tries to write the Yukawa couplings of the quark sector, he introduces $2\times (3\times 3)$ complex Yukawa coupling constants (or 36 real couplings) from which there result ten observable parameters (six masses and four angles). Even allowing the unobservable phase degrees of freedom of the quarks (twelve left and right handed quark phases minus the baryon phase), fifteen redundant parameters are left with. Going beyond the simple data fitting in the SM, to have any predictive result(s) from the Yukawa coupling structure the number of coupling parameters should be drastically reduced. Symmetries are used to reduce the number of couplings. The early attempt toward this direction has been suggested by Weinberg such that the mixing angles are related to some ratios of quark masses \cite{Wein79mass}. Because of the numerical coincidence of $\sin\theta_C\simeq \sqrt{m_d/m_s}$, this approach attracted a great deal of attention \cite{Fritzsch79} and constituted the most fruitful Yukawa textures until recently.

On the other hand, the mixing angles are not very close to zero, in particular for the case of the neutrino mixing angles. The $\nu_\mu-\nu_\tau$ mixing angle $\theta_{\mu\tau}$, being close to $45^{\rm o}$ which is called {\it bi-maximal}, is not imagined to arise from some kind of a mass ratio. A naive guess to obtain this large mixing angle is from relating some Yukawa couplings to be identical. The simplest such idea is to employ a permutation symmetry $S_3$ to have a bi-maximal form \cite{Harrison95}. In fact, the permutation symmetry $S_3$ has been discussed as early as in 1970s \cite{Perm70s,SegWel79}. Initiated by Harrison, Perkins and Scott \cite{Harrison02}, the permutation symmetry $S_4$ and its subgroup $A_4$ allowing triplet representations have been extensively used to obtain {\it tri-bimaximal} PMNS matrix \cite{S4A4}. This idea is generalized to consider more discrete symmetries for quark and lepton mixing angles \cite{MoreDSym,Review10}. For the quark mixing angles, the knowledge on electroweak scale physics is enough. But for the neutrino masses, one needs more information beyond the SM spectrum. If one does not introduce any singlet neutrinos at the electroweak scale, the neutrino masses appear as dimension five operators which need the information at a high energy scale. To have any predictive results, {\it screening of the Dirac flavor structure} have been used toward this end, for example in \cite{KimPark,Lindner05,Keum06}.

In this paper, we introduce the {\it dodeca-symmetry} $D_{12}$ to obtain the quark and lepton mixing angles. For the $D_{n}$ symmetry, there already exists a nice paper by Blum, Hagedorn and Lindner (BHL), trying to obtain the Cabibbo angle of $\theta_C\simeq 13^{\rm o}$ \cite{Blum08}. Here, we do not attempt to obtain the exact Cabibbo angle observed near $ 13^{\rm o}$, but we try to obtain $\theta_C= 15^{\rm o}$ from the $D_{12}$ symmetry.
The bi-dodeca form PMNS matrix we obtain here is
\dis{
 V_{\rm PMNS} =
  \left ( \begin{array}{ccc}
   \mathrm{cos} \frac{\pi}{6} & \mathrm{sin} \frac{\pi}{6} & 0  \\
   - \frac{1}{\sqrt{2}} \mathrm{sin} \frac{\pi}{6} & \frac{1}{\sqrt{2}} \mathrm{cos} \frac{\pi}{6} & -\frac{1}{\sqrt{2}}  \\
   - \frac{1}{\sqrt{2}} \mathrm{sin} \frac{\pi}{6} & \frac{1}{\sqrt{2}} \mathrm{cos} \frac{\pi}{6} & \frac{1}{\sqrt{2}}
  \end{array} \right)
}
which looks as simple as the tri-bimaximal form.
The shift of $\theta_C$ from $15^{\rm o}$  to $13.14^{\rm o}$ may be achieved by terms breaking the $D_{12}$ symmetry and/or its renormalization from a grand unification (GUT) scale down to the electroweak scale.

The above developments can be summarized as follows. If quark mixing angles are small, the ideas for obtaining mixing angles, for example $\theta_{\alpha\beta}$ of $\cos\theta_{\alpha\beta}=|V^{CKM}_{\alpha\beta}|$, can be symbolically written as two functions $f$ and $g$,
\dis{
\sin\theta_{\alpha\beta} \simeq f_{\alpha\beta}\left(\theta_i,\{\frac{\lambda_a}{\lambda_b}\}\right)+ g_{\alpha\beta}\left(\{\frac{m_k}{m_l}\},
\{\frac{\lambda_a}{\lambda_b}\}\right)\label{eq:mixgeneral}
}
where $\theta_i$ are the angles arising from discrete symmetries, $\{\frac{m_k}{m_l}\}$ is a set of ratios of complex masses (whose magnitudes, \ie the Yukawa couplings, are defined to be less than 1), and $\{\frac{\lambda_a}{\lambda_b}\}$ is a set of ratios of complex parameters (whose magnitudes are defined to be not greater than 1) in the Higgs potential. Weinberg's calculation corresponds to $f_{\alpha\beta}=0$ \cite{Wein79mass}, and Pakvasa and Sugawara's calculation includes $f_{\alpha\beta}$ with the assumption that the VEVs of Higgs fields obtain complex phases due to a finite range of coupling constants in the Higgs potential \cite{Perm70s}. The BHL attempt corresponds to $g_{\alpha\beta}=0$  and $f_{\alpha\beta}( \theta_i ,\{\frac{\lambda_a}{\lambda_b}\}) =f_{\alpha\beta}(\theta_i)$. Our study of mixing angles in this paper follows the spirit of BHL.

When one employs discrete symmetries for a $3\times 3$ matrix, it is required that the three mass eigenvalues are different to fit to the three observed masses as the one we show below as $x, w, z$ for $D_{12}$ or three real numbers $a,b,c$ (the diagonal one plus two off-diagonal ones) of Ref. \cite{Harrison02} for the cyclic permutaions of $S_3$. The anticipated finite mixing angles from discrete symmetries must appear from the phases for the case of $Z_N$ symmetries or from a similar unit one complex number such as the cube root of unity for the cyclic permutaions of $S_3$.

However, in explaining the mixing angles through the nontrivial phases of the Higgs VEVs, there exists an important problem to be resolved. Usually, many Higgs fields are used in this kind of attempts, and hence the Higgs potential can be very complicated. Then, it is not clear whether the desired phase choices are allowed from the Higgs potential without fine-tuning of parameters. The BHL case seems to use a fine-tuning. Here, we attempt to resolve this problem by assigning several Higgs fields to different $D_{12}$ representations, and in addition introducing more symmetries such as the Peccei-Quinn (PQ) symmetry and/or a $Z_2$ symmetry.

It is desirable that the needed discrete symmetry arises from an ultraviolet completion of the model. One may consider an ultraviolet completion of a global symmetry also. Along this line, some discrete symmetries and an approximate PQ symmetry
 have been considered in the $Z_{6-II}$ \cite{Kobayashi08}, $Z_3$ \cite{Kim88} and $Z_{12-I}$ \cite{ChoiKimKim} orbifold compactifications of the heterotic string. In this kind of ultraviolet completion, we need to know all the particles in the theory and their interactions to find out the approximate global and discrete symmetries. Even if we know all the particle content, a general study of discrete symmetries, which needs some identical strength of couplings from an ultraviolet completed theory, is limited because the coupling constants involve geometrical factors. On the other hand, for global symmetries a mere knowledge on the existence (but not the strength) of the terms is the requirement \cite{ChoiKimKim}. For example, the Yukawa coupling between three fields each located at three different fixed points involve a geometrical factor $e^{-cA}$ where $A$ is the area made by the three points, and hence requiring identical couplings is farfetched for considering fields located at  numerous sets of three different fixed points. But if three fields appear at the same fixed point, then there certainly exists a discrete symmetry in their Yukawa coupling structure \cite{Kim05mu}. At the string unification or the grand unification scale, which will be simply called the GUT scale, there appear numerous SM singlets. In this paper, we study the dodeca-symmetry at the field theory level assuming the existence of these numerous SM singlets at the GUT scale.

In Sec. \ref{sec:Model}, we present a dodeca-model for quark and lepton masses and mixing angles. We specify a vacuum, leading to $\theta_C=15^{\rm o}, \theta_{\rm sol}=30^{\rm o}$, and $\theta_{\mu\tau}=45^{\rm o}$. Here, we also comment on the flavor changing neutral couplings (FCNC) due to radial Higgs fields. In Sec. \ref{sec:SpBreaking}, we study a model for the spontaneous symmetry breaking of the dodeca-symmetry, leading to the vacuum of Sec. \ref{sec:Model}. In Sec. \ref{sec:higherorder}, we calculate the next order generation of mixing angles to obtain the shift of $\theta_C$ and generation of $\theta^{\rm CKM}_{23},\theta^{\rm CKM}_{32}$, etc. Sec. \ref{sec:Conclusion} is a conclusion. In Appendix, we list several formulae of the $D_{2N}$ symmetry which are used in the text.

\section{Model}\label{sec:Model}

Four key observations about the mixing angles are \cite{PData08},
\begin{enumerate}
\item The Cabibbo angle, determined by the (11) element of the CKM matrix, is $\theta_{C}\simeq 13.14^{\rm o}$.
\item The $\nu_\mu-\nu_\tau$ mixing is maximal, i.e. close to $\theta_{\mu\tau}\simeq 45^{\rm o}$.
\item The (11) element of the PMNS matrix determined by the Solar and KamLand experiments \cite{Kayser08}, for which $\theta_{\rm sol}$ will be called `Solar angle', is consistent with $\theta_{\rm sol}=30^{{\tiny +9.5~}\rm }_{\tiny -5.4}$ degrees.
\item The (13) element of the PMNS matrix is almost zero, and the (13) element of the CKM matrix is very small \cite{Wolfenstein}.
\end{enumerate}
A corollary of Items 1 and 3 is the famous observation on the Solar angle and the Cabibbo angle \cite{KimPark}
\dis{
\theta_{\rm sol}+\theta_{C}\simeq 45^{\rm o}.
}
Observation of 2 and 3 has led to many discrete symmetry models on neutrino masses, typically riding on the bandwagon of the tri-bimaximal form \cite{S4A4}. But tri-bimaximal form does not exactly lead to $\theta_{\rm sol}=30^{\rm o}$ but to a value $35.3^{\rm o}$, both of which are consistent with the data \cite{Kayser08}.

In this paper, we look for a discrete symmetry allowing the leading value to $\theta_{\rm sol}=30^{\rm o}$. Assuming the leading values of $\theta_C, \theta_{\rm sol}$, and $\theta_{\mu\tau}$ as $15^{\rm o},30^{\rm o}$, and $45^{\rm o}$, respectively, $30^{\rm o}$ is a key angle, implying an integer $12=\frac{360^{\rm o}}{30^{\rm o}}$. Since there also appears the half of $30^{\rm o}$, in addition we consider $Z_2$. Thus, we may consider $Z_{12}\times Z_2$. This structure arises from the dihedral group $D_{12}$.
Thus, we employ the dihedral symmetry $D_{12}$.

\subsection{Higgs representations}\label{subsec:Higgs}

The standard model (SM) fermions obtain Dirac masses, coupling to the Higgs doublets. At the high energy scale, there can exist numerous Higgs singlets.
Some Higgs doublets with specified quantum numbers can be composites of a Higgs doublet and singlet(s).

The Higgs doublets giving mass to up type quarks are supplied with superscript $u$, the Higgs doublets giving mass to down type quarks are supplied with superscript $d$, and the Higgs doublets giving mass to leptons are supplied with superscript $l$. The Higgs doublets form the following doublets and singlets under $ D_{12} $,

\dis{
H_{0}^{u} : \textbf{1}_{++}, ~\left ( \begin{array}{c}   H_{1}^{\prime u}  \\
   H_{2}^{\prime u}  \end{array} \right) : \textbf{2}_{1}, ~\left ( \begin{array}{c}   H_{1}^{\prime  \prime u}  \\
   H_{2}^{\prime \prime u}  \end{array} \right) : \textbf{2}_{3}\label{rep:Hu}
}
\dis{
H_{0}^{d} : \textbf{1}_{++}, ~H_{0}^{\prime d} : \textbf{1}_{++},~
   \left ( \begin{array}{c}   H_{1}^{\prime d}  \\   H_{2}^{\prime d}
  \end{array} \right) : \textbf{2}_{2}\label{rep:Hd}
}
\dis{
H_{0}^{l} : \textbf{1}_{++}, ~H_{0}^{\prime l} : \textbf{1}_{++},~
   \left ( \begin{array}{c}   H_{1}^{\prime l}  \\   H_{2}^{\prime l}
  \end{array} \right) : \textbf{2}_{2}\label{rep:Hl}
}
where subscripts denote the kinds of $D_{12}$ (five) doublet and (four) singlet representations discussed in Appendix.
For $H^{l}$s to couple to leptons but not to quarks and for $H_0^{d}$  to couple to quarks but not to leptons, we can introduce a leptonic $Z_3$ discrete symmetry such that charged singlet leptons, lepton doublets and $H^{l}$s carry  $Z_3$ quantum number 1 and all the other fields, except the singlet neutrinos, carry  $Z_3$ quantum number 0. [The  $Z_3$ quantum number of singlet neutrinos will be commented in Subsec. \ref{subsubsec:neutrino}.] Moreover, we need to prevent the possibility that $H^u$s enter in the down quark terms and $H^d$s in the up quark terms. If supersymmetry is imposed, such a mixing is forbidden since $H^u$s and $H^d$s have different U(1)$_Y$  quantum numbers. However, we do not consider supersymmetry here, and it is possible that $H^{u \dagger}$s contribute to the $d-$quark sector. To avoid this possibility, we impose a U(1) PQ symmetry. The PQ symmetry also plays an important role in solving the strong CP problem and in forbidding the unwanted terms in the Higgs potential to make our vacuum phase choice reasonable. The PQ charge assignment and Higgs potential are discussed in Sec. \ref{sec:SpBreaking}.  For example, suppose we impose PQ charge +1 to both $H_{0}^{u}$ and $H_{0}^{d}$. This prevents a combination  $H_{0}^{\dagger u}+H_{0}^{d}$. If we are to explain Yukawa coupling in the context of Froggatt-Nielsen scheme \cite{Froggaatt79}, the PQ charge could be differently imposed from the ones we presented here. In this scenario, the Higgs doublets can be thought of as some products of fields of a doublet field and singlet fields carrying PQ charges. But, here we do not delve into this detail.

Note that we have not introduced the following Higgs which mix the $D_{12}$ doublet and singlet fermions:
\dis{
\left ( \begin{array}{c}
   H_{1}^{u}  \\   H_{2}^{u}  \end{array} \right) : \textbf{2}_{1},~
   \left ( \begin{array}{c}   H_{1}^{d}  \\   H_{2}^{d}
  \end{array} \right) : \textbf{2}_{1},~
   \left ( \begin{array}{c}   H_{1}^{l}  \\   H_{2}^{l}
  \end{array} \right) : \textbf{2}_{1}.\label{rep:Hufun}
  }
Even though we write some couplings with the fields of (\ref{rep:Hufun}) below, we will eventually set those entries zero, either by not introducing the lowest order $D_{12}$ representations as above or by assuming their vanishing VEVs.

Since we assign a few Higgs doublets for the same charge fermions (up-type quarks, down-type quarks, and charged leptons), there exist the FCNCs from neutral Higgs fields \cite{GlWein77}. In the unitary gauge, the Higgs doublets are represented as
\dis{
H_I = e^{2i(\tau_3 P^0_I+\tau^+ P^-_I +\tau^- P^+_I)/V_I}
\left( \begin{array}{c}\frac{1}{\sqrt2}(V_I+\rho^{0}_I) \\ \rho_I^{-} \end{array}\right)\label{eq:HiggsUnitary}
}
where $\tau_i$ are the SU(2) generators for the doublet, and $\rho_I$ and $P_I$ are the radial and phase fields of a complex Higgs field $H_I$: $H_I^0=\{\rho^{0}_I,P^{0}_I\}$ and $H_I^-=\{\rho^{-}_I,P^{-}_I\}$ with $P^+=P^{-*}$.

There also appear numerous SM singlets at the GUT scale which can form $D_{12}$ doublets and singlets. These SM singlets, denoted as $S, S^\prime$ and $\Phi$ fields, are defined at the appropriate places where they are explicitly needed, for example in Subsubsec. \ref{subsubsec:neutrino}  on the neutrino masses and Sec. \ref{sec:SpBreaking} on the spontaneous symmetry breaking.

In this Section, we will present appropriate vacuum expectation values (VEVs) of the Higgs fields consistent with the $D_{12}$ symmetry. These VEVs are chosen such that successful mixing angles result. It is equivalent to choosing a specific vacuum out of degenerate vacua, and hence there appears the cosmological domain wall problem in the standard Big Bang cosmology, which is assumed to be resolved by passing through an inflationary epoch.
 \cite{Riva}

\subsection{The quark sector}

We represent SU(2)$_W$ quark doublets as the upper case $Q$'s and SU(2)$_W$ quark singlets as the lower case $q$'s. Thus, the left-handed SM quarks are
\dis{
 Q_1 = \left ( \begin{array}{c}   u  \\   d
  \end{array} \right),&
  ~u^c, ~ d^c,~~  Q_2 = \left ( \begin{array}{c}   c  \\   s
  \end{array} \right),
  ~c^c,~ s^c\\
    Q_3 &= \left ( \begin{array}{c}  t  \\   b
  \end{array} \right),
 ~t^c,~ b^c
}
Three SM quark doublets form a doublet and a singlet under $ D_{12} $,
\dis{  \left ( \begin{array}{c}
   Q_1  \\   Q_2  \end{array} \right) :
  \textbf{2}_{1}, ~~  Q_3 :   \textbf{1}_{++}
}
where subscripts denote the kinds of $D_{12}$ representations out of five  $D_{12}$ doublets and four $D_{12}$ singlets.
Six SM quark singlets form two doublets and two singlets under $ D_{12} $,
\dis{  \left ( \begin{array}{c}
   u^c  \\   c^c  \end{array} \right) :  \textbf{2}_{2}, ~~
  t^c :    \textbf{1}_{++},   ~~   \left ( \begin{array}{c}
   d^c  \\   s^c  \end{array} \right) :  \textbf{2}_{1},  ~~
  b^c :    \textbf{1}_{++}
}
where subscripts denote the kinds of $D_{12}$ representations.

\subsubsection{The up quark Yukawa couplings}
The tensor product of
$Q_3(\textbf{1}_{++}) \times t^{c}(\textbf{1}_{++})$ implies that it can couple to $H^{u} _0(\textbf{1}_{++})$, leading to the coupling, viz. Eq. (\ref{eq:dbdbtosinga}),
\dis{
y_{1}^{u} H_{0}^{u} \bar{t}_L t_R
}
where $y_1^u$ is the Yukawa coupling constant.

On the other hand, since $\mathbf{2}_2$ Higgs does not exist,
\dis{
Q_3(\textbf{1}_{++}) \times
 \left ( \begin{array}{c}   u^{c}  \\   c^{c}
  \end{array} \right)(\textbf{2}_{2} )  \nonumber
  }
cannot make $D_{12}$ singlet, but
\dis{
 \left ( \begin{array}{c}  Q_1  \\   Q_2
  \end{array} \right)(\textbf{2}_{1} )  \times t^c (\textbf{1}_{++})\nonumber
  }
can couple to
\dis{
 \left ( \begin{array}{c}
   H_{1}^{u}  \\   H_{2}^{u}  \end{array} \right)(\textbf{2}_{1} ).
 \nonumber\\
}
So, we consider the coupling
\dis{ y_{3}^{u} (H_{2}^{u} \bar{u}_L t_R + H_{1}^{u} \bar{c}_L t_R )
}
where we used Eq. (\ref{eq:dbdbtosinga}).
Consideration of
\dis{
 \left ( \begin{array}{c}
   Q_1  \\   Q_2  \end{array} \right)( \textbf{2}_{1} )  \times
 \left ( \begin{array}{c}
   u^{c}  \\  c^{c}
  \end{array} \right) ( \textbf{2}_{2} )\nonumber
  }
allows its coupling, via Eq. (\ref{eq:dbdbtosinga}),  to
\dis{
 \left ( \begin{array}{c}   H_1^{\prime u}  \\   H_2^{\prime u}
  \end{array} \right)(\textbf{2}_{1})
   ~ \mathrm{and} ~
   \left ( \begin{array}{c}   H_1^{\prime \prime u}  \\   H_2^{\prime \prime u}
  \end{array} \right)(\textbf{2}_{3}),\nonumber
 }
i.e. the following Yukawa coupling
 \dis{y_{4}^{u} (  H^{ \prime u}_{1} \bar{u}_L c_R +  H^{ \prime u}_{2} \bar{c}_L u_R ) + y_{5}^{u} (  H^{\prime \prime u}_{2} \bar{u}_L u_R +  H^{\prime \prime u}_{1} \bar{c}_L c_R ).
}

These couplings are summarized by the following up mass matrix
\dis{
 M^{(u)}=
  \left ( \begin{array}{ccc}
   y_5^u H_2^{\prime \prime u} & y_4^u H_1^{\prime u} & y_3^u H_2^u  \\
   y_4^u H_2^{\prime u} & y_5^u H^{\prime \prime u}_1 & y_3^u H_1^u  \\
   0 & 0 & y_1^u H_0^u
  \end{array} \right)
}

One can construct a desirable mixing matrix by taking the zero VEV of $ (H_1^u , H_2^u)^T $, which represents $ ( \textbf{2} _{1} - \textbf{1} _{++} ) $ quark mixing if not vanished.

One may also think of it as  $ (H_1^u , H_2^u)^T $ Higgs is forbidden by some kinds of symmetry. That means,  $ \textbf{1} _{++} $ and $ \textbf{2} _{1} $ quarks are completely separated.

\vskip 0.5cm
\centerline{\bf The FCNC problem}

Since we introduced more than one Higgs VEV to the up-type quark masses, in general there exists the FCNC problem among up-type quarks \cite{GlWein77}. Using Eq. (\ref{eq:HiggsUnitary}), the up-type quarks have the following cubic couplings,

\begin{widetext}

\dis{
&\sum_i \overline{q}_L^{i}\sum_j \sum_I f^{(u)I}_{ij} e^{2i\theta_{ij}^I} e^{2i(\tau_3 P^0_I+\tau^+ P^-_I +\tau^- P^+_I)/V_I}
\left( \begin{array}{c}\frac{1}{\sqrt2}(V_I+\rho^{(u)0}_I) \\ \rho_I^{(u)-} \end{array}\right)u_R^j +{\rm h.c.} \\
&= \left\{
\begin{array}{l}
  \overline{q}_L M u_R \leftarrow \frac{V}{\sqrt2} (\overline{u}_L,\overline{d}_L)\sum_I f^{(u)I}_{ij} e^{2i\theta_{ij}^I} \cos\alpha_I e^{2i(\tau_3 P^0_I+\tau^+ P^-_I +\tau^- P^+_I)/V_I}
  \left( \begin{array}{c}u_R^j \\ 0 \end{array}\right) +{\rm h.c.} \\
   \overline{q}_LP u_R \leftarrow \frac{V}{\sqrt2}(\overline{u}_L,\overline{d}_L)\sum_I f^{(u)I}_{ij}  e^{2i\theta_{ij}^I} {\cos\alpha_I} e^{2i(\tau_3 P^0_I+\tau^+ P^-_I +\tau^- P^+_I)/V_I}
  \left( \begin{array}{c} u_R^j \\ 0 \end{array}\right)+{\rm h.c.} \\
   \overline{q}_L\rho^{u0} u_R \leftarrow \frac{1}{\sqrt2} (\overline{u}_L,\overline{d}_L)\sum_I f^{(u)I}_{ij} e^{2i\theta_{ij}^I}
  \left( \begin{array}{c} \rho^{(u)0}_I u_R^j \\ \rho_I^{(u)-}  u_R^j  \end{array}\right) +{\rm h.c.}
\end{array}  \right. \label{eq:SU2coup}
}

\end{widetext}
where the complexity of the Yukawa coupling is denoted as $ e^{2i\theta_{ij}^I}$ and $\cos\alpha_I=|V_I|/V$ with $V= \sqrt{\sum_I |V_I|^2}$.
The phase fields $P$'s do not contribute to the cubic Yukawa couplings since they are rotated away when we diagonalize the mass terms. The FCNC problem exist through the neutral $\rho^{(u)0}_I$ couplings to $\overline{u}c+\overline{c}u$. The FCNC problem can be removed either by assuming almost degenerate radial Higgs fields or superheavy Higgs fields. However, superheavy mass is not desirable where $D_{12}$ symmetry breaking occurs at the electroweak scale.
We will comment more on this later in the down quark sector which gives the most stringent bound on the FCNC.

\vskip 0.5cm
\centerline{\bf $D_{12}$ breaking}
The $ D_{12} $ symmetry is broken down to a smaller symmetry generated by $ b $,  by assigning the VEVs as
\dis{
 \left ( \begin{array}{c}
    H_1^u  \\ H_2^{u}
  \end{array} \right) ( \textbf{2}_{1} ) = v_u \left ( \begin{array}{c} 1  \\ 1
  \end{array} \right),
  \left ( \begin{array}{c}
    H_1^u  \\ H_2^{u}
  \end{array} \right) ( \textbf{2}_{2} ) = v^{\prime}_u \left ( \begin{array}{c} 1  \\ 1
  \end{array} \right),
 \\
   y_4^u  \left ( \begin{array}{c}  H^{\prime u}_1  \\
   H^{\prime u}_2\end{array} \right) (\textbf{2}_{1} ) = w_u
     \left ( \begin{array}{c} 1  \\ 1
  \end{array} \right),\\
   y_5^u  \left ( \begin{array}{c}  H^{\prime \prime u}_1  \\
   H^{\prime \prime u}_2\end{array} \right) (\textbf{2}_{3} ) = z_u
     \left ( \begin{array}{c} 1  \\ 1
  \end{array} \right),
\\ y_1^u H_0^u = x_u .\label{eq:VEVupH}
}
The exact breaking pattern can be found in Appendix, or in Ref. \cite{Blum08}.
Not introducing Eq. (\ref{rep:Hufun}) is equivalent to setting
$v_u=0$ and $v_u^{\prime}=0$ in the mass matrix,
and we consider only $ \textbf{2}_{2} $ vacuum and $ D_{12} $ is then broken down to $ D_2 $ generated by $  a^6 $ and $ ba^6 $, where $a$ and $b$ are generators of $D_{12}$ defined in Appendix.
 Thus, the mass matrix
becomes
 \begin{equation}
 M^{(u)}=
  \left ( \begin{array}{ccc}
   w_u & z_u & 0  \\
   z_u & w_u & 0  \\
   0 & 0 & x_u
  \end{array} \right)\label{eq:upmasses}
\end{equation}
which is diagonalized by the following unitary matrix,
\dis{
 U_u =
  \left ( \begin{array}{ccc}
   \frac{1}{\sqrt{2}} & \frac{1}{\sqrt{2}} & 0  \\
   - \frac{1}{\sqrt{2}} & \frac{1}{\sqrt{2}} & 0  \\
   0 & 0 & 1
  \end{array} \right)
}
Then, the mass eigenvalues appear as
\dis{
\tilde M^{(u)2} &= U^{\dagger}_u (M^{(u)}  M^{(u)\dagger}_u) U_u\\
 &=
  \left ( \begin{array}{ccc}
   (w_u -z_u )^2 & 0 & 0  \\
   0 & (w_u +z_u)^2 & 0  \\
   0 & 0 & x_u^2
  \end{array} \right)
}
which allow three independent mass values for the $u,c$, and $t$ quarks.

\subsubsection{The down quark Yukawa couplings}
Calculating the down type quark Yukawa couplings in the same way, we obtain
\dis{
 M^{(d)} =
  \left ( \begin{array}{ccc}
   y_5^d H_2^{\prime d}  & y_4^d H_0^{\prime d}  & y_3^d H_2^d  \\
   y_4^d H_0^{\prime d}  & y_5^d H_1^{\prime d}  & y_3^d H_1^d  \\
   y_2^d H_2^d & y_2^d H_1^d & y_1^d H_0^d
  \end{array} \right)
}

The $D_{12}$ symmetry is broken down to a $ D_2 $ generated by $ba$ and $ a^6$, by assigning VEVs (for $v_d=0$) as
\dis{
   \left ( \begin{array}{c}  H_1^d  \\ H_2^{d}
  \end{array} \right) ( \textbf{2}_{1} )
  &= v_d  \left ( \begin{array}{c} \mathrm{e}^{-i \phi}  \\ 1
  \end{array} \right) ,\\
  ~~
   y_5^d  \left ( \begin{array}{c} H^{\prime d}_1   \\ H^{\prime d}_2
  \end{array} \right) &( \textbf{2}_{2} )
  = w_d  \left ( \begin{array}{c} \mathrm{e}^{-2i \phi} \\ 1
  \end{array} \right) ,\\
   y_1^d H_0^d &= x_d ,~~y_4^d H^{\prime d}_0=z_d\label{eq:VEVdownH}
}
where we choose $ \phi = \frac{2 \pi}{12} $, the smallest angle with the dodeca-symmetry.
Not introducing Eq. (\ref{rep:Hufun}) is equivalent to setting $v=0$ in the mass matrix, and we obtain the following $d$ quark mass matrix,
\dis{
 M^{(d)} =
  \left ( \begin{array}{ccc}
   w_d & z_d & 0  \\
   z_d & w_d \mathrm{e} ^{-2 i \phi} & 0  \\
   0 & 0 & x_d
  \end{array} \right)\label{eq:downmasses}
}
which is diagonalized by the unitary matrix
\dis{
 U_d =
  \left ( \begin{array}{ccc}
   \frac{1}{\sqrt{2}} & \frac{1}{\sqrt{2}} \mathrm{e}^{i \phi} & 0  \\
   - \frac{1}{\sqrt{2}} \mathrm{e}^{-i \phi} & \frac{1}{\sqrt{2}} & 0  \\
   0 & 0 & 1
  \end{array} \right).
}
Then, the diagonalized mass matrix squared becomes
\dis{
& \tilde M^{(d)2} = U^{\dagger}_d (M^{(d)} M^{ (d) \dagger}) U_d\\
 &=
  \left (\begin{array}{ccc}
   w_d^2 + z_d^2 - 2w_d z_d \cos \phi & 0 & 0  \\
   0 & w_d^2 + z_d^2 + 2w_d z_d \cos \phi  & 0  \\
   0 & 0 & x_d^2
  \end{array} \right)
}

Since we introduced more than one Higgs VEV to the down-type quark masses through $H^d$, the FCNC problem exists also among the down-type quarks for which the $K_L-K_S$ mass difference gives the most stringent bound. For the tree level effective interaction through the radial Higgs $\rho^{(d)0}_I, (f^2/M_\rho^2)\bar d s \bar d s +\rm h.c.$,
the $K_L-K_S$ mass difference due to $\rho^{(d)0}$ exchange is estimated as
\cite{BrancoCP99},
\dis{
\Delta m_K& \simeq 2\frac{f^2}{M^2_{\rho^{(d)0}}} B \frac{f_K^2 m_K}{12}\left(-1+\frac{m_K^2}{(m_s+m_d)^2}\right) \\
&\simeq  4 \frac{f^2}{M^2_{\rho^{(d)0}}}  m_K f_K^2 B
 \label{eq:KLKSdiff}}
where $B$ is the bag parameter in the range $\frac13-1$ \cite{GuptaR98}. Eq. (\ref{eq:KLKSdiff}) can be compared to the experimental value of $(3.483 \pm 0.066) \times 10^{-15}$ GeV \cite{PData08}. For $f\simeq 10^{-3}$ from $m_s/100~\rm GeV$, we obtain $M_{\rho^{(d)0}} >2.2\sqrt{B}(f_K/f_\pi)$ TeV. If two degenerate radial Higgs of mass of order 250 GeV contribute to the $\Delta S=2$ effective interaction with the opposite signs, then we obtain
\dis{
\Delta m_K = \frac{4f^2}{\overline{M}_{\rho}^{\,4}}\Delta M^2 m_K f_K^2
}
which leads to  $\sqrt{\Delta M^2} \lesssim 26[\sqrt{B}(f_K/f_\pi)]^{-1}$ GeV. Thus, it seems that the FCNC problem can be resolved with a reasonable range of Yukawa couplings and radial Higgs masses.

\subsubsection{ The CKM matrix}\label{subsubsec:CKM}
The CKM mixing matrix becomes
\dis{
 V_{\rm CKM} = U^{\dagger}_u  U_d
 =
  \left ( \begin{array}{ccc}
   \mathrm{e}^{-i \phi /2} \mathrm{cos} \frac{\phi}{2} &
   i \mathrm{e}^{i \phi /2}  \mathrm{sin} \frac{\phi}{2}  & 0  \\
   i \mathrm{e}^{-i \phi /2}  \mathrm{sin} \frac{\phi}{2}
    & \mathrm{e}^{i \phi /2} \mathrm{cos} \frac{\phi}{2}   & 0  \\
   0 & 0 & 1
  \end{array} \right)
}
Note that  the (11) element of $V_{CKM}$ gives the Cabibbo angle
$\theta_C=  \frac{\phi}{2} = 15^{\circ} $.

Note that Arg.Det.$M_q$ from Eqs. (\ref{eq:upmasses}) and (\ref{eq:downmasses}) does not give a vanishing value. Therefore, this model if stopped here has the strong CP problem \cite{KimRMP10}, and therefore we need to introduce a very light axion.
On the other hand, Segr\'e and Weldon  introduced a calculable $\bar\theta$ model with an $S_3$ permutation symmetry such that Arg.Det.$M_q$ is 0 at tree level and  remains zero up to one loop level \cite{SegWel79}. Even if \cite{SegWel79} discusses the strong CP problem with a discrete symmetry, our model does not belong to this category and moreover does not belong to any discrete symmetry model asserting Arg. Det. $M_q=0$ at tree level.

\subsection{The lepton sector}
The SM leptons are assigned as
\dis{
 &L_1 = \left ( \begin{array}{c} \nu _e  \\ e^-
  \end{array} \right), ~e^c,\\ & L_2 = \left ( \begin{array}{c}
   \nu _{\mu}  \\ \mu ^- \end{array} \right), ~\mu ^c,
  ~~
   L_3 = \left ( \begin{array}{c} \nu _{\tau}  \\ \tau ^-
  \end{array} \right),~\tau ^c
}
Moreover, we introduce two kinds of right handed neutrinos, $ (n_1, n_2, n_3) $ and $ ( N_1 , N_2 , N_3 ) $, as suggested in Ref. \cite{KimPark}. In Ref. \cite{KimPark}, the double seesaw mechanism has been employed to screen the Dirac flavor structure in the neutrino mass matrix, and hence the light-neutrino mass matrix becomes directly proportional to a heavy-neutrino ($n$) mass matrix.\footnote{There can be other methods to hide the light-neutrino Dirac masses as tried in \cite{Lindner05}.}
 For such a screening to occur, the coupling between $L_i$ and $N_i$ should be proportional to the coupling between $n_i$ and $N_i$. With the following renormalizable Yukawa couplings
\dis{
f_{IJ}^{(lN)}N^I H^{\nu N} L^J + f_{IJ}^{(Nn)}N^I n^J S^{nN} + f_{IJ}^{(nn)}n^I n^J S^n, \label{cpl}
}
we can specify the required relations. The required condition is $f_{IJ}^{(lN)} \propto f_{IJ}^{(Nn)}$. Such an (almost) exact proportionality could arise in the context of GUT. Suppose $L_i$ and $n_i$ belong to the same multiplet of a larger gauge group, say, $F_1$, and $H^{\nu N}$ and $S^{nN}$ belong to the same multiplet, say $S$. Let $F_2$ be the multiplet to which $N$ neutrinos belong. Then both $f_{IJ}^{(lN)}N^I H^{\nu N} L^J$ and $f_{IJ}^{(Nn)}N^I n^J S^{nN}$ come from the same interaction, $S F_1 F_2$, with a common coupling constant. If the see-saw scale is at the high energy scale so that the splitting of couplings are not so large, then $f_{IJ}^{(lN)}$ is almost the same as $f_{IJ}^{(Nn)}$.

 For example, in the SU(6) GUT model \cite{SU6}, one of right handed neutrino ($n$ in this case) and lepton doublet belong to the same representation, say $\mathbf{\overline{6}}^M $, another right handed neutrino $N$ is an SU(6) singlet and $S$ belongs to $\mathbf{6}^S$ representation. Then, the first two terms in Eq. (\ref{cpl}) have the same origin, $f (\bar{\mathbf{6}}^M N \mathbf{6}^S) $. When SU(6) is broken down to SU(5)$\times$U(1), splitting of the coupling $f$ into $f^{lN}$ and $f^{Nn}$ occurs, at the order of $\frac{f^2}{16 \pi^2} \mathrm{ln}(\frac{M_{see~saw}}{M_{GUT}})$. Supposing $M_{see~saw} \sim 10^{14}$ GeV , $M_{GUT} \sim 10^{16}$ GeV, and $f \sim O(1)$ then the splitting effect is about $0.03$, \ie only 3 per cent.

On the other hand, we can also construct a term $\mathbf{15}^M \bar{\mathbf{6}}^M \bar{\mathbf{6}}^H$ to form the Yukawa coupling. As splitting $\bar{\mathbf{6}}^M \rightarrow \bar{\mathbf{5}}^M + n$ occurs, we obtain various terms where $n$ couples to the SM matter as well as to the as-yet-unobserved massive particles. Since thye Yukawa couling of the SM particles (in the SU(5) language, $y(\mathbf{10}^M \bar{\mathbf{5}}^M \bar{\mathbf{5}}^H)$) should be present, it might be hard to prevent all these terms toward the screening in the double see-saw mechanism  \cite{KimPark}.  But even in this case, the coupling $y$ could be much smaller than $f$ since $y< O(10^{-2})$, and the screening effects in double see-saw mechanism is a very good approximation. For example, the $\tau$ lepton mass is about 1.8 GeV at electroweak scale and therefore its Yukawa coupling is about $10^{-2}$. Since the RG equation of each Yukawa coupling is proportional to the Yukawa coupling itself, we expect that the correction from unified Yukawa coupling is small, $\frac{y^2}{16 \pi^2} \mathrm{ln}(\frac{M_{EW}}{M_{GUT}})\sim O(10^{-2}-10^{-3})$, which means that $y$ is still much smaller than the $O(1)$ coupling $f$ even at the GUT scale.

We give the following $ D_{12}$ assignments for the SM leptons,
\dis{
    L_1 :     \textbf{1}_{++},  ~~
  \left ( \begin{array}{c} L_2  \\ L_3
  \end{array} \right) :\textbf{2}_{1}\\ e^c :  \textbf{1}_{++},  ~~
  \left ( \begin{array}{c} \mu ^c  \\ \tau ^c \end{array} \right) :
  \textbf{2}_{1}
}
For the heavy-neutrinos whose mass matrix is proportional to the light-neutrino mass matrix, we assign
\dis{
   \left ( \begin{array}{c} n_1 + i n_2  \\ n_1 - i n_2
  \end{array} \right) : \textbf{2}_{2}  ~~ n_3 : \textbf{1}_{++}~ .
}
Note that we combined two Majorana neutrinos to make a complex field required for a doublet representation of $D_{12}$.

We need not specify the representation content of $ N_i $ if it applies to the double see-saw mechanism \cite{KimPark}.

\subsubsection{Charged leptons}

For charged lepton masses, we use the Higgs doublets presented in Eq. (\ref{rep:Hd}). Then, the mass matrix of charged leptons is given by
\dis{ M^{(l)} =
  \left ( \begin{array}{ccc}
   y_1^l H_0^l & y_2^l H_2^l & y_2^l H_1^l  \\
   y_3^l H_2^l & y_5^l H^{\prime l}_2  & y_4^l H^{\prime l}_0 \\
   y_3^l H_1^l & y_4^l H^{\prime l}_0 & y_5^l H^{\prime l}_1
  \end{array} \right)
}

The  $ D_{12} $ symmetry is broken down to $ D_2 $, generated by $a^6$ and $ ba^6 $, by assigning the VEVs as
\dis{
   \left ( \begin{array}{c}  H_1^l  \\ H_2^l
  \end{array} \right) (\textbf{2}_{1}) =& v_l
     \left ( \begin{array}{c}-1  \\    1\end{array} \right) ,~~
   y_5^l  \left ( \begin{array}{c} H^{\prime l}_1  \\ H^{\prime l}_2
  \end{array} \right)( \textbf{2}_{2}) = w_l \left ( \begin{array}{c}
    1  \\  1 \end{array} \right),\\
 & y_1^l H_0^l = x_l ,~~y_4^l H^{\prime l}_0 =z_l.
}
Note that we introduced $H^l$'s which are different from $H^d$'s.
Not introducing Eq. (\ref{rep:Hufun}) is equivalent to setting $v=0$ in the mass matrix, and the $ \textbf{1} _{++} $ lepton and the $ \textbf{2} _1 $ leptons are not mixed,
\dis{
 M^{(l)} =
  \left ( \begin{array}{ccc}
   x_l & 0 & 0  \\
   0 & w_l & z_l  \\
   0 & z_l & w_l
  \end{array} \right).
}
The charged lepton mass squared, $ M_l M^{\dagger}_l $, is diagonalized by
\dis{
 U_l =
  \left ( \begin{array}{ccc}
   1 & 0 & 0  \\
   0 & \frac{1}{\sqrt{2}} &  \frac{1}{\sqrt{2}}  \\
   0 & -  \frac{1}{\sqrt{2}} &  \frac{1}{\sqrt{2}}
  \end{array} \right).
}

Since we introduced more than one Higgs VEV to the charged lepton masses, the FCNC problem exists among the charged leptons, e.g. for the $\mu\to  e^- e^- e^+$ decay. For an effective interaction of $\mu\to e^- e^- e^+$ decay, $(f^2/M_{\rho}^2) \bar{e} e  \bar{e} \mu$, the decay rate is estimated as
\dis{
\Gamma(\mu \to e^- e^- e^+) = \frac{1}{8}\frac{f^4}{M_{\rho}^4} \frac{m_{\mu}^5}{192 \pi^3}
}
which must lead to the branching ratio less than $10^{-12}$ \cite{PData08}. This requires  $M_{\rho} > 190$ GeV for $f \sim 10^{-3}$.

\subsubsection{Neutrinos}\label{subsubsec:neutrino}
In models with the screening of the Dirac flavor structure in the neutrino mass matrix, the light neutrino mass matrix is assumed to be proportional to the heavy $n$ neutrino mass matrix, $M^{(\nu)}\propto M^{(n)}$. So the number of heavy Majorana neutrinos $n$ is the same as that of the SM doublet neutrinos $\nu$. The SM singlet neutrinos $n$ are required to obtain masses by the VEVs of SM singlet Higgs fields $S$. So, the dodeca-symmetry of the needed SM singlet Higgs fields $S$ is
\dis{
&\quad S_{0}^{n} : \textbf{1}_{++}, ~~S^{\prime n}_{0} : \textbf{1}_{++},\\
&   \left ( \begin{array}{c}
   S^{n}_{1}  \\
   S^{n}_{2}
  \end{array} \right) : \textbf{2}_{1},
  \left ( \begin{array}{c}
   S^{\prime n}_{1} \\
  S^{\prime n}_{2}
  \end{array} \right) : \textbf{2}_{4}
}
If we try to complete the theory at high energy, we may need to consider the $N$-type neutrinos and more Higgs fields, singlets
$ S^{nN} $ and doublets $ H^{\nu N} $, to allow the $n-N$ and $ \nu - N $ mixing for the double seesaw mechanism.

To forbid $S$ to couple to charged leptons or quarks, we need to assign $Z_3$ quantum number as stated. Therefore, $S$ and $n$ neutrinos have $Z_3$ quantum number $-1$.

Now, the neutrino mass  matrix can be written as

\begin{widetext}

\dis{
 M^{(\nu)} =
  \left ( \begin{array}{ccc}
   y_4^n 2S^{\prime n}_{0} + y_5^n (S^{\prime n}_1  +S^{\prime n}_2) &
    iy_5^n (S^{\prime n}_2 - S^{\prime n}_1) &
    y_3^n (S_1^n + S_2^n )  \\
   i y_5^n (S^{\prime n}_2 - S^{\prime n}_1) &
    y_4^n 2S^{\prime n}_0- y_5^n (S^{\prime n}_1 + S^{\prime n}_2) &
    i y_3^n (S_2^n - S_1^n )  \\
   y_2^n ( S_2^n + S_1^n ) &
   i y_2^n (S_2^n - S_1^n) &
    y_1^n S_0^n
  \end{array} \right)
}
\end{widetext}

We require that the $D_{12}$ symmetry is broken down to $D_2 $ generated
by $a^3 $ and $ ba $ (for $v_n=0$)
\dis{
&   \left ( \begin{array}{c}
    S_1^n  \\
   S_2^{n}
  \end{array} \right) (\textbf{2}_{1})
  = v_n
     \left ( \begin{array}{c}
     \mathrm{e}^{-i \phi /2}  \\
     \mathrm{e}^{i \phi /2}
  \end{array} \right),\\
 &\quad   y_5^n \left ( \begin{array}{c}
  S^{\prime n}_1   \\
  S^{\prime n}_2
  \end{array} \right) (\textbf{2}_{4} )
  = w_n
     \left ( \begin{array}{c}
    \mathrm{e}^{-i \phi} \\
    \mathrm{e}^{i \phi }
  \end{array} \right),\\
&\quad y_1^n S_0^n = x_n,~~y_4^n S^{\prime n}_0 =z_n
}
where $ \phi = \frac{2 \pi}{12} \times 2 $ . Also, taking $ v=0 $, we obtain
\dis{
 M^{(\nu)} =\left ( \begin{array}{ccc}
   2(z_n+w_n \mathrm{cos} \phi ) & -2w_n \mathrm{sin} \phi & 0  \\
   -2w_n \mathrm{sin} \phi & 2(z_n-w_n \mathrm{cos} \phi ) & 0  \\
   0 & 0 & x_n
  \end{array} \right)
}
which is diagonalized by
\dis{
 U_{\nu} =
  \left ( \begin{array}{ccc}
   \mathrm{cos} \frac{\phi}{2} & \mathrm{sin} \frac{\phi}{2} & 0  \\
   - \mathrm{sin} \frac{\phi}{2} & \mathrm{cos} \frac{\phi}{2} & 0  \\
   0 & 0 & 1
  \end{array} \right)
}

  \begin{equation}
\tilde M^{(\nu)} = U_{\nu}^{\dagger} M^{(\nu)} U_{\nu}
 =\left ( \begin{array}{ccc}
   2(z_n+w_n) & 0 & 0  \\
   0 & 2(z_n-w_n)  & 0  \\
   0 & 0 & x_n
  \end{array} \right).
\end{equation}
The three independent neutrino masses can be fitted to the observed neutrino mass ratios from the neutrino oscillation data.

\subsubsection{The PMNS matrix}
Now the PMNS matrix is calculated as
\dis{
 V_{\rm PMNS} = U_l^{\dagger}U_{\nu} =
  \left ( \begin{array}{ccc}
   \mathrm{cos} \frac{\pi}{6} & \mathrm{sin} \frac{\pi}{6} & 0  \\
   - \frac{1}{\sqrt{2}} \mathrm{sin} \frac{\pi}{6} & \frac{1}{\sqrt{2}} \mathrm{cos} \frac{\pi}{6} & -\frac{1}{\sqrt{2}}  \\
   - \frac{1}{\sqrt{2}} \mathrm{sin} \frac{\pi}{6} & \frac{1}{\sqrt{2}} \mathrm{cos} \frac{\pi}{6} & \frac{1}{\sqrt{2}}
  \end{array} \right).
}
which is the desired bi-dodeca mixing form. The third column represents the bi-maximal mixing, and the other angles are multiples of $30^{\rm o}$, which is the dodeca mixing.

\section{Spontaneous breaking of $D_{12}$}\label{sec:SpBreaking}

The vacuum choices of Sec. \ref{sec:Model} for desired quark and lepton mixing angles must be consistent with the Higgs potential.  Couplings between Higgs and their complex conjugates are restricted by $SU(2)_L \times U(1)_Y\times U(1)_\Gamma\times Z_3\times D_{12}$ where $U(1)_\Gamma$ is the PQ symmetry and $Z_3$ is the leptonic one discussed below Eq. (\ref{rep:Hl}).  For example, by the $U(1)_Y$ symmetry, $H_u H_d$ and $(H_uH_d^{\dagger})(H_u^{\dagger}H_d)$ are allowed, whereas $(H_uH_d^{\dagger})^2$ is forbidden. In this section, we study how Higgs potential could be minimized and suggest what other symmetry is needed toward the vacuum choice of Sec. \ref{sec:Model}.

In Higgs potential, the most problematic terms are those containing $D_{12}$ doublets $H^{\prime d}$, $S^n$, and $S^{\prime n}$, which have non-trivial phases so that we have to verify whether our phase choice is not spoiled. By imposing another symmetry such as the PQ symmetry or a $Z_2$ symmetry, we can forbid the unwanted terms. We show how this possibility is realized for $D_{12}$ doublets. The potential containing $D_{12}$ singlets can be treated in the same way.

For $D_{12}$ doublets $H$s and $H^\prime$s, we need to know the tensor products which can be found in Appendix.

For example, consider the tree level Higgs potential made of $D_{12}$ doublets. For the quartic tensor products, the following terms are allowed,
\dis{ (H^{\prime u}H^{\prime u})(H^{\prime d}H^{\prime d}),~~  (H^{\prime u}H^{\prime u})(H^{\prime u \dagger}H^{\prime u \dagger})
\\
 (H^{\prime u}H^{\prime u})(H^{\prime u \dagger}H^{\prime d}),~~  (H^{\prime d}H^{\prime d})(H^{\prime d \dagger}H^{\prime d \dagger})
\\
 (H^{\prime d}H^{\prime d})(H^{\prime d \dagger}H^{\prime u}),~~  (H^{\prime u}H^{\prime d})(H^{\prime u \dagger}H^{\prime d \dagger})
 \\
  (H^{\prime u}H^{\prime d})(H^{\prime u \dagger}H^{\prime u}),~~  (H^{\prime u}H^{\prime d})(H^{\prime d \dagger}H^{\prime d})
  \\
   (H^{\prime d \dagger}H^{\prime u})(H^{\prime u \dagger}H^{\prime d}),~~  (H^{\prime u}H^{\prime d})(H^{\prime u }H^{\prime d })
   \\
    (H^{\prime u \dagger}H^{\prime u})(H^{\prime u \dagger}H^{\prime u}),~~  (H^{\prime d \dagger}H^{\prime d})(H^{\prime d \dagger}H^{\prime d})
    \\
  (H^{\prime u \dagger}H^{\prime u})(H^{\prime d \dagger}H^{\prime d})
   }
and their Hermitian conjugates.
 Suppose we introduce the PQ charge +1 to both $H^{\prime u}$ and $H^{\prime d}$.
$H^{\prime u}$ might be replaced by $H^{\prime \prime u}$, but in this case the term such as $(H^{\prime \dagger}_u H^{\prime \prime}_u)(H^{\prime \dagger}_d H^{\prime}_d) + h.c.$ do not minimize our vacuum phase choice. For both $H^{\prime u}$ and $H^{\prime \prime u}$ not to appear in the same tree level quartic terms, we assign different PQ charges to $H^{\prime u}$ and $H^{\prime \prime u}$.
Then, the following terms survive,
    \dis{  (H^{\prime u}H^{\prime u})(H^{\prime u \dagger}H^{\prime u \dagger}), ~~  (H^{\prime d}H^{\prime d})(H^{\prime d \dagger}H^{\prime d \dagger})
\\
 (H^{\prime u}H^{\prime d})(H^{\prime u \dagger}H^{\prime d \dagger}),~~
   (H^{\prime d \dagger}H^{\prime u})(H^{\prime u \dagger}H^{\prime d})
   \\
    (H^{\prime u \dagger}H^{\prime u})(H^{\prime u \dagger}H^{\prime u}),~~  (H^{\prime d \dagger}H^{\prime d})(H^{\prime d \dagger}H^{\prime d})
    \\
  (H^{\prime u \dagger}H^{\prime u})(H^{\prime d \dagger}H^{\prime d})
   }
and terms with $H^{\prime u}$ replaced by $H^{\prime \prime u}$.
The Lagrangian contains the following terms,

   \dis{\vert H_{1}^{\prime u} \vert ^2 \vert H_{2}^{\prime u} \vert ^2 , ~~
   \vert H_{1}^{\prime u} \vert ^4 + \vert H_{2}^{\prime u} \vert ^4
   \\
   \vert H_{1}^{\prime \prime u} \vert ^2 \vert H_{2}^{\prime \prime u} \vert ^2 , ~~
   \vert H_{1}^{\prime \prime u} \vert ^4 + \vert H_{2}^{\prime \prime u} \vert ^4
   \\
   \vert H_{1}^{\prime d} \vert ^2 \vert H_{2}^{\prime d} \vert ^2 , ~~
   \vert H_{1}^{\prime d} \vert ^4 + \vert H_{2}^{\prime d} \vert ^4
   \\
   (\vert H_{1}^{\prime u} \vert ^2 + \vert H_{2}^{\prime u} \vert ^2 )^2, ~~ (\vert H_{1}^{\prime u} \vert ^2 - \vert H_{2}^{\prime u} \vert ^2 )^2
   \\
   (\vert H_{1}^{\prime d} \vert ^2 + \vert H_{2}^{\prime d} \vert ^2 )^2, ~~ (\vert H_{1}^{\prime d} \vert ^2 - \vert H_{2}^{\prime d} \vert ^2 )^2
  \\
    ( H_{1}^{\prime u} H_{1}^{\prime d}) ( H_{1}^{\prime u \dagger} H_{1}^{\prime d \dagger})+  ( H_{2}^{\prime u} H_{2}^{\prime d}) ( H_{2}^{\prime u \dagger} H_{2}^{\prime d \dagger})
\\
    ( H_{2}^{\prime u} H_{1}^{\prime d}) ( H_{2}^{\prime u \dagger} H_{1}^{\prime d \dagger})+  ( H_{1}^{\prime u} H_{2}^{\prime d}) ( H_{1}^{\prime u \dagger} H_{2}^{\prime d \dagger})
  \\
    ( H_{1}^{\prime \prime u} H_{1}^{\prime  d}) ( H_{1}^{\prime \prime u \dagger} H_{1}^{\prime  d \dagger})+  ( H_{2}^{\prime \prime u} H_{2}^{\prime  d}) ( H_{2}^{\prime \prime u \dagger} H_{2}^{\prime  d \dagger})
\\
    ( H_{2}^{\prime \prime u} H_{1}^{\prime  d}) ( H_{2}^{\prime \prime u \dagger} H_{1}^{\prime  d \dagger})+  ( H_{1}^{\prime \prime u} H_{2}^{\prime  d}) ( H_{1}^{\prime \prime u \dagger} H_{2}^{\prime  d \dagger})
  \\
    ( H_{2}^{\prime d \dagger} H_{1}^{\prime u}) ( H_{1}^{\prime u \dagger} H_{2}^{\prime d })+  ( H_{1}^{\prime d \dagger} H_{2}^{\prime u}) ( H_{2}^{\prime u \dagger} H_{1}^{\prime d })
\\
    ( H_{2}^{\prime d \dagger} H_{2}^{\prime u}) ( H_{2}^{\prime u \dagger} H_{2}^{\prime d })+  ( H_{1}^{\prime d \dagger} H_{1}^{\prime u}) ( H_{1}^{\prime u \dagger} H_{1}^{\prime d })
  \\
    ( H_{2}^{ \prime d \dagger} H_{1}^{\prime \prime u}) ( H_{1}^{\prime \prime u \dagger} H_{2}^{\prime d })+  ( H_{1}^{\prime \prime d \dagger} H_{2}^{\prime \prime u}) ( H_{2}^{\prime \prime u \dagger} H_{1}^{\prime  d })
\\
    ( H_{2}^{\prime  d \dagger} H_{2}^{\prime \prime u}) ( H_{2}^{\prime \prime u \dagger} H_{2}^{ \prime d })+  ( H_{1}^{\prime d \dagger} H_{1}^{\prime \prime u}) ( H_{1}^{\prime u \dagger} H_{1}^{\prime d })
\\
(\vert  H_{1}^{\prime u} \vert ^2 + \vert  H_{2}^{\prime u} \vert ^2 )(\vert  H_{1}^{ \prime d} \vert ^2 + \vert  H_{2}^{ \prime d} \vert ^2 )
\\
(\vert  H_{1}^{\prime \prime u} \vert ^2 + \vert  H_{2}^{\prime \prime u} \vert ^2 )(\vert  H_{1}^{ \prime  d} \vert ^2 + \vert  H_{2}^{\prime  d} \vert ^2 )
\\
( H_{2}^{\prime \prime u \dagger} H_{1}^{\prime \prime u})^2 + ( H_{1}^{\prime \prime u \dagger} H_{2}^{\prime \prime u})^2
 \label{eq:Higgpoten}
}
Our phase choice of VEVs in Subsec. \ref{subsec:Higgs}  must be consistent with the above potential. To investigate it in more detail, we pay attention to the last two terms. The other terms are not introducing phases. Let $\delta_1$ and $\delta_2$ be phases of $H_{1}^{\prime \prime u}$ and $H_{2}^{\prime \prime u}$, respectively. For Hermiticity and $D_{12}$ invariance, the coupling constant should be real. The last term depends on phases through
\dis{
\cos(2(\delta_1-\delta_2))
}
and our vacuum choice $\delta_1=\delta_2=0$ minimize it provided the coupling constant is negative. It is worth to note here that, if at least one of two $D_{12}$ Higgs doublets were in the same representation, it is very hard to minimize the potential toward the desired vacuum property. For example, suppose that both $H^{\prime u}$ and $H^{\prime d}$ are in the same representation. In this case, the following terms are allowed.
\dis{ (H_{1}^{\prime u}H_{2}^{\prime d})(H_{2}^{\prime u \dagger}H_{1}^{\prime d \dagger})  +h.c .
}
For the invariance under the generator $b$ of $D_{12}$, the overall coefficient must be real. Let $\alpha^u_1, ~\alpha^u_2, ~\alpha^d_1, ~\alpha^d_2$ be the phases of Higgs VEV of $H_{1}^{\prime u}, H_{2}^{\prime u}, H_{1}^{\prime d }$,   and $H_{2}^{\prime d }$ , respectively. So, this quartic term has the phase dependence $\cos(\alpha^u_1-\alpha^u_2-\alpha^d_1+\alpha^d_2)$ and our vacuum choice does not minimize it.

The quadratic terms allowed by gauge and PQ symmetries are, viz. Eq. (\ref{rep:Hu}),
  \dis{H^{\prime u \dagger} H^{\prime u},\quad H^{\prime \prime u \dagger} H^{\prime \prime u}, \quad H^{\prime d \dagger} H^{\prime d}
  }
  and their Hermitian conjugates. $D_{12}$ singlets are
   \dis{\vert H^{\prime u }_1 \vert ^2 + \vert H^{\prime u}_2 \vert ^2
  ,\\
  \vert H^{\prime \prime u }_1 \vert ^2 + \vert H^{\prime \prime u}_2 \vert ^2
\\
   \vert H^{\prime d }_1 \vert ^2 + \vert H^{\prime d}_2 \vert ^2~.
  }
These quadratic terms may introduce negative mass squared toward achieving the VEVs of neutral members of the Higgs doublets.

The forbidden terms at tree level can appear integrating out heavy fields whose VEVs possibly break the assumed symmetries. These could be used to explain the vacuum choice of $H^{\prime d }$ and therefore explains how $D_{12}$ can be the flavor symmetry. For example, consider the quartic terms made of $D_{12}$ doublet Higgs without conjugate (or starred) fields. Then, we have
\dis{ \textbf{1}_{++} :~
 &  (H^{\prime u}_2 H^{\prime d}_1) (H^{\prime u}_1 H^{\prime d}_2),~ (H^{\prime u }_1 H^{\prime d }_1)( H^{\prime u }_2 H^{\prime d }_2)
\label{eq:onepp}
 }
\dis{ \textbf{1}_{+-} :~
 (H^{\prime u}_1 H^{\prime d}_1 )^2 + (H^{\prime u}_2 H^{\prime d}_2)^2
 }
\dis{ \textbf{1}_{-+} :~
 (H^{\prime u}_1 H^{\prime d}_1 )^2 - (H^{\prime u}_2 H^{\prime d}_2)^2
 }
 \dis{ \textbf{2}_2 :~
 & \left ( \begin{array}{c}  (H^{\prime u}_2 H^{\prime d}_1)^2  \\
   (H^{\prime u}_1 H^{\prime d}_2)^2
  \end{array} \right)
  \\
 ~& \left ( \begin{array}{c}  (H^{\prime u}_1 H^{\prime d}_2)(H^{\prime u}_1 H^{\prime d}_1)  \\
(H^{\prime u}_2 H^{\prime d}_1)(H^{\prime u}_2 H^{\prime d}_2)
  \end{array} \right)
 }
 \dis{ \textbf{2}_4 :
  \left ( \begin{array}{c} ( H^{\prime u}_2 H^{\prime d}_1 )(H^{\prime u}_1 H^{\prime d}_1)  \\
 ( H^{\prime u}_1 H^{\prime d}_2 )(H^{\prime u}_2 H^{\prime d}_2)
  \end{array} \right)
 }
 Note that the term given in Eq. (\ref{eq:onepp}) is forbidden by the PQ symmetry of Table \ref{tab:PQcharges}.

Let us introduce a $D_{12}$ doublet ${\bf 2}_4$ which is denoted as a SM singlet scalar $\Phi$,
\dis{ \Phi =  \left ( \begin{array}{c} \Phi_1  \\ \Phi_2
  \end{array} \right) : \textbf{2}_4~ \label{eq:PhiDefine}
  }
Using $\Phi$, the allowed quartic couplings are obtained. In addition, we note

\begin{itemize}
\item The dimension-5 $D_{12}$ allowed couplings are
  \dis{
  &  \lambda [\Phi^{\dagger}_1 ( H^{\prime u}_2  H^{\prime d}_1)( H^{\prime u}_1  H^{\prime d}_1) +
  \Phi^{\dagger}_2 ( H^{\prime u}_1  H^{\prime d}_2)( H^{\prime u}_2  H^{\prime d}_2)]
 \label{eq:allfive}
   }
\item The dimension-6 $D_{12}$ allowed couplings are
 \dis{
& \zeta_1 \Phi^{\dagger}_1 \Phi^{\dagger}_2 (H^{\prime u}_2 H^{\prime d}_1)( H^{\prime u}_1 H^{\prime d}_2)+ \zeta_2 \Phi^{\dagger}_1 \Phi^{\dagger}_2  (H^{\prime u}_1 H^{\prime d}_1)( H^{\prime u}_2 H^{\prime d}_2)\\
&  + \zeta_3 [ \Phi^{\dagger 2}_2(H^{\prime u}_2 H^{\prime d}_1)( H^{\prime u}_1 H^{\prime d}_1)+ \Phi^{\dagger 2}_1 (H^{\prime u}_1 H^{\prime d}_2)( H^{\prime u}_2 H^{\prime d}_2)].\label{eq:allsix}
 }
 \end{itemize}
Here, $ \textbf{2}_8 $ is shown to be equivalent to $ \textbf{2}_4 $ by applying a $D_{12}$ transformation $b$ of Eq. (\ref{eq:btrans})
\dis{
  \left ( \begin{array}{cc} 0 & 1  \\  1 & 0
  \end{array} \right)
    \left ( \begin{array}{c} x_1  \\ x_2
  \end{array} \right) (\textbf{2}_8) ~~:~~ \textbf{2}_4 .
}
Operators with dimension more than 7 are highly suppressed and hence they can be ignored. All effective quartic terms coupling to $\Phi$ do not give the vacuum discussed in Sec. \ref{sec:Model}. The terms except those discussed in Sec. \ref{sec:Model} must be forbidden by some symmetry or at least highly suppressed. For example, if we choose the VEV of $\Phi$ as $\frac{1}{\sqrt2}(\rm{exp}(-i 2 \pi /3), 1)^T$, only the dimension-5 operator is independent of the phase choices given in Eqs. (\ref{eq:VEVupH}) and (\ref{eq:VEVdownH}).  However, this vacuum choice at this stage is dangerous as discussed before. To forbid Eq. (\ref{eq:allfive}), we introduce a $Z_2$ symmetry:  $\Phi\to -\Phi$. With this discrete symmetry, a dimension-6 operator of the form
\dis{
(\Phi_1^3 + \Phi_2^3)(\Phi_1^{\dagger 3} + \Phi_2^{\dagger 3})\label{eq:Dimsix}
}
is not forbidden. Moreover, this term favors the direction which makes $\langle \Phi_1^3  \rangle + \langle \Phi_2^3 \rangle = 0 $.  With this dimension 6 potential, our vacuum choice is not the minimum.

 Since dimension-5 operators are forbidden, we may choose an alternate direction $\Phi \propto \frac{1}{\sqrt2}( 1,\rm{exp}(-i  \pi /3))^T$. Then, our vacuum choice of Sec. \ref{sec:Model} corrsponds to the minimum.

 The fact that $\Phi$ has a VEV with phase could affect the phase of Yukawa coupling very much by the higher order corrections. This correction is tiny, of order of $10^{-3}$, possibly contributing to the Cabibbo angle correction from 15$^{\rm o}$ to a smaller one. Due to the symmetries of our model, the largest contribution is $\Phi^{\dagger} \Phi = \Phi_1^{\dagger} \Phi_1 + \Phi_2^{\dagger} \Phi_2 $, which is independent of the phase; therefore it does not affect the mixing angle at all. The correction to Yukawa coupling can be written as
$ H^{d \prime} Q d^c \{1+ \frac{\Phi^{\dagger} \Phi}{M^2}+\rm{(dimension~~8~~operators)}\}$.
The effects of dimension 7 operator is of order $10^{-3}$ compared to the tree level value. Moreover, with SUSY even $\Phi^{\dagger} \Phi$ does not appear by holomorphy and $Z_2$ symmetry, and hence the $\Phi$ effects is even smaller.

Since Arg.Det.$M_q$ is nonzero at the required vacuum
as commented in Subsubsec. \ref{subsubsec:CKM}, we need a PQ symmetry
to have a strong CP solution. Since our
discussion on the dodeca-symmetry is at the electroweak
scale, the PQ symmetry U(1)$_\Gamma$ must be manifest at a
high energy scale of the axion window. To confine to
the axion window, the model-independent axion \cite{Witten84} may
not be useful as commented for example in \cite{KimNilles09}. So, with
the electroweak dodeca-symmetry the very light axions with the decay constant in the axion window of $10^{10-12}$ GeV may be a possibility toward the strong CP solution \cite{hadronic,DFSZ}.

\subsection{A detailed high energy model}\label{subsec:ultracompl}
Therefore, to forbid dimension-5 operator, consider the following $ D_{12} $ representations, and in addition the $ U(1)_{\Gamma}$ PQ charges and $Z_2$ assignments  of Table \ref{tab:PQcharges}.
\begin{table}[h]
\begin{center}
\begin{tabular}{c|ccccc}
Fields& $H^{\prime u}$ &~$H^{\prime \prime u}$ &~$H^{\prime d}$&~$X$ &~$\Phi$ \\
\hline
 $D_{12}$& ${\bf 2}_1$&~${\bf 2}_3$&~${\bf 2}_2$&~ ${\bf 2}_3$&~ ${\bf 2}_4$
 \\
 ${\Gamma}$&$+1$& $+2$ & $+1$ & $-2$& $+2$
 \\
 $Z_2$& + & + & + & + & --
 \end{tabular}
 \end{center}
 \caption{The PQ charges of some Higgs fields}\label{tab:PQcharges}
\end{table}

\begin{figure}[t!]
 \begin{center}
 \epsfig{figure=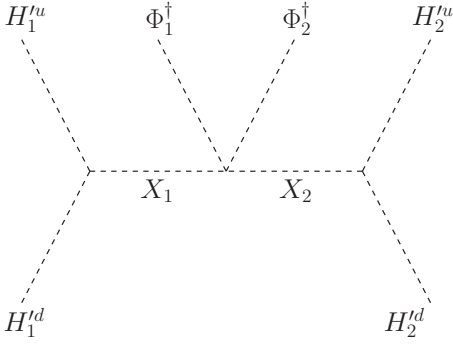,width=6cm,height=4.5cm,angle=0}
\caption{ The Feynman diagram leading to the $\zeta$ term of (\ref{eq:allsix}).  }
\label{fig1:Feyn}
\end{center}
\end{figure}

Suppose $X$ is much heavier than the Higgs $H'$ and $\Phi$. By the U(1)$_{\Gamma} $ and the $Z_2$ invariance, the only allowed renormalizable interactions except for the self interactions or Higgs-Higgs interactions (which are responsible for determining the magnitude of VEVs) are
\dis{
 X H^{\prime u} H^{\prime d}~~ \mathrm{and}~~ \Phi^2 X^2~.
 }
In terms of the component fields, these become
\dis{
X_1 H^{\prime u}_2 H^{\prime d}_2 +X_2 H^{\prime u}_1 H^{\prime d}_1
~~\mathrm{and}~~
X_1X_2\Phi_1\Phi_2 .
}
In this case, the following effective interactions are allowed at the tree level,
\dis{
\Phi^{\dagger}_1 \Phi^{\dagger}_2 (H^{\prime u}_1 H^{\prime d}_1)(H^{\prime u}_2 H^{\prime d}_2).
}
which are shown in Fig. \ref{fig1:Feyn}.

In this model, choosing  the VEV of $ \Phi$ as $ ( 1, \mathrm{e}^{-i \pi /3})^T $,
 we obtain the desired vacuum allowing the $\zeta$ term of (\ref{eq:allsix}).  Note that $Z_2$ plays an important role in prohibiting the unwanted terms, e.g. all terms in (\ref{eq:allfive}).

\section{Violation of $D_{12}$: Shift of $\theta_C$ and generation of small angles}\label{sec:higherorder}
\begin{figure}[t!]
 \begin{center}
 \epsfig{figure=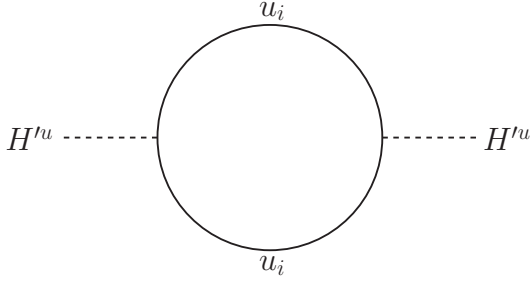,width=7cm,height=3.7cm,angle=0}
\caption{ The Feynman diagrams leading to the violation of the dodeca-symmetry of the $H^{\prime u}$ doublet through loops for $i=1$ or 2. Here, $u=u_1$ and $c=u_2$. }
\label{fig2:Loop}
\end{center}
\end{figure}

The violation of the $D_{12}$ symmetry can arise from a few sources.
Firstly, the $D_{12}$ symmetry can be broken by a disparity in the masses within a $D_{12}$ doublet as shown in Fig. \ref{fig2:Loop}. Second, some explicit $D_{12}$ symmetry breaking terms such as $\lambda_i(i=1,\cdots,4)$ terms of Eq. (\ref{eq:allfive}) can be present in the Lagrangian.

In this section, we estimate the magnitude of $D_{12}$ breaking by Fig.  \ref{fig2:Loop} and then introduce phenomenologically needed  $D_{12}$ breaking terms to fill the vanishing entries  of the leading $V_{\rm CKM}$ and $V_{\rm PMNS}$ and to shift $\theta_C$. Next we argue how these $D_{12}$ breaking terms can arise from a more fundamental principle.
With this scheme, we argue that the shift of $\theta_C$ and $V^{\rm CKM}_{23,32}$ are of the same order while $V^{\rm CKM}_{13,31}$ need to be a bit more suppressed, and then estimate the magnitude of $V^{\rm PMNS}_{13}$.

For the quantum correction, different quark masses cause the symmetry breaking. Consider, for example, the up quark sector. The Higgs VEVs, and equivalently the Higgs masses are corrected by fermion loops. Among Higgs fields in our model, $ H^{\prime u}_0$ just scales $z$ and does not affect the diagonalizing unitary matrix. $H^{\prime u}_1$ couples only to one fermion, and so does $ H^{\prime u}_2 $. Their loop corrections makes the $w$ entries of mass matrix elements (11) and (22) different. As done in the electroweak SU(2)$_W$ breaking by the disparity of $t$ and $b$ quark masses \cite{Veltman77}, the fermion one-loop correction is given by,
 \dis{
 \delta m_{Higgs}^2 = - \frac{N_c \vert y \vert ^2}{8 \pi^2}
 \left[ \Lambda^2 +m^2 - 3 m^2 \mathrm{ln}(\frac{\Lambda^2 }{m^2}) \right]
 }
where $m$ is the mass of the fermion in the loop of Fig. \ref{fig2:Loop}.  It corresponds to the $c$ quark for $ H^{\prime u}_1$ and to the $u$ quark for $H^{\prime u}_2 $. Since each quark mass is very small compared to the Higgs VEVs, such corrections do not affect the change of the mixing angles very much. Comparing the Higgs VEVs and the fermion masses, with $ \Lambda \sim v_{Higgs}$, we estimate

 \dis{
& \frac{w_{(22)}^2 - w_{(11)}^2}{ w^2 }
 \sim \frac{N_c }{ v^2_{Higgs} }
 \Big[-(m_c^2 - m_u^2)
 \\
&+ 3\left(m_c^2 \mathrm{ln}(\frac{\Lambda^2 }{m_c^2})- m_u^2 \mathrm{ln}(\frac{\Lambda^2 }{m_u^2})\right)\Big] < 0.002
 }
 which is very small.

Therefore, we have to consider the explicit breaking terms, to shift $\theta_C$ by 2$^{\rm o}$. In general, a unitary $ 3 \times 3 $ matrix can be represented by three Euler angles and a phase,
\dis{\left (\begin{array}{ccc}
   1 & 0 & 0  \\
   0 & c_{23} &  s_{23} \\
   0 & -s_{23} & c_{23}
  \end{array} \right)
  \left (\begin{array}{ccc}
   c_{13} & 0 & s_{13} \mathrm{e}^{-i \delta}  \\
   0 & 1 &  0 \\
   -s_{13} \mathrm{e}^{i \delta} & 0 & c_{13}
  \end{array} \right)
   \left (\begin{array}{ccc}
   c_{12} & s_{12} & 0  \\
   -s_{12} & c_{12} &  0 \\
   0 & 0 & 1
  \end{array} \right)
   }

For the CKM angles, $ \theta^{\rm CKM}_{23} $ is of order 0.04 which must be generated at the next level.
   If $ U_u$ is given by
   \dis{
  & U_u=
   \\
  &   \left (\begin{array}{ccc}
   \frac{1}{\sqrt2} & \frac{1}{\sqrt2} & 0  \\
  -\frac{1}{\sqrt2} & \frac{1}{\sqrt2} & 0 \\
   0 & 0 & 1
  \end{array} \right)
  \left (\begin{array}{cc}
   2 \times 2~ {\rm unitary} & 0  \\
      {\rm  matrix} & 0 \\
   0\quad \quad 0  & 1
  \end{array} \right)
  \left (\begin{array}{ccc}
   1 & 0 & 0  \\
  0 & c_{23} & -s_{23} \\
   0 & s_{23} & c_{23}
  \end{array} \right)
   }
From $ V_{CKM} = U_u^{\dagger} U_{d} $, the third matrix gives $ \theta_{23}$ and the second matrix gives a correction to the Cabibbo angle.

For the lepton sector,
\dis{
  U_l
  =\left (\begin{array}{ccc}
   1 &0 & 0 \\
 0 & \frac{1}{\sqrt2} &  \frac{1}{\sqrt2} \\
   0 & -\frac{1}{\sqrt2} & \frac{1}{\sqrt2}
  \end{array} \right)
  \left (\begin{array}{ccc}
   c_{13} &0 & s_{13} \\
 0 & 1 &  0 \\
   - s_{13} & 0 & c_{13}
  \end{array} \right)
  }
gives a nonzero $ \theta_{13}$ from $ V_{PMNS} = U_l^{\dagger} U_{\nu} $.

Let us present these corrections with the following explicit $D_{12}$ breaking terms,
\dis{
& i[\epsilon_1 y_4^u H^{\prime u}_0 \bar{u}_L c_R\\
  &+
 \epsilon_2 (y_1^u H^u_0 - y_4^u H^{\prime u}_0 - y_5^u H^{\prime u}_2)
 (\bar{u}_L + \bar{c}_L)t_R ] + h. c.\label{eq:Dvioqu}
 }
 Then,  we obtain
\begin{widetext}
\dis{
 M^{(u)} =
  \left (\begin{array}{ccc}
   w_u & z_u(1+i \epsilon_1) & i \epsilon_2 (x_u-z_u-w_u)  \\
   z_u(1-i \epsilon_1) & w_u  &  i \epsilon_2 (x_u-z_u-w_u) \\
   -i \epsilon_2 (x_u-z_u-w_u) & -i \epsilon_2 (x_u-z_u-w_u) & x_u
  \end{array} \right)
}
which can be diagonalized by
\dis{U_u =\left (\begin{array}{ccc}
   \frac{1}{\sqrt2} & \frac{1}{\sqrt2} & 0 \\
   - \frac{1}{\sqrt2} & \frac{1}{\sqrt2} & 0   \\  0 & 0 & 1
  \end{array} \right) \left (\begin{array}{ccc}  -i& \frac{\epsilon_1}{2}  & 0 \\
    \frac{\epsilon_1}{2} & -i & 0   \\ 0 & 0 & 1 \end{array} \right)
   \left (\begin{array}{ccc}  1 & 0 & 0 \\ 0 & 1 & - \sqrt2 \epsilon_2  \\ 0 &  \sqrt2 \epsilon_2 & 1
  \end{array} \right)=
 \left (\begin{array}{ccc}
 \frac{1}{\sqrt2}(-i+\frac{\epsilon_1}{2}) &\frac{1}{\sqrt2}(-i+\frac{\epsilon_1}{2})  & i\epsilon_2 \\
 \frac{1}{\sqrt2}(i+\frac{\epsilon_1}{2})  &\frac{1}{\sqrt2}(-i-\frac{\epsilon_1}{2})
 &  i\epsilon_2   \\  0 & \sqrt2  \epsilon_2 & 1
\end{array} \right) +O( \epsilon^2_{1,2}).
 }
For  ${\epsilon_1}\simeq 0.07$, the shift of $\theta_C$ is about $ \frac{\epsilon_1}{2} = 0.035 \sim \mathrm{sin} 2^{\rm o }$, it gives rise to the needed correction of the Cabibbo angle: $15^{\rm o}\to 13^{\rm o}$.
 For $ \sqrt2 {\epsilon_2}\simeq 0.04$, $\vert V_{cb} \vert $ is obtained at the observed value  $4.12 \times 10^{-2}$ \cite{PData08}.
The entire form of the CKM matrix is
 \dis{V_{\rm CKM} =
 \left (\begin{array}{ccc}
   i \mathrm{e}^{-i \phi/2} \cos x & - \mathrm{e}^{i \phi/2} \sin x & 0 \\
   - \mathrm{e}^{-i \phi/2} \sin x & i \mathrm{e}^{i \phi/2} \cos x & \sqrt2\epsilon_2  \\
    \sqrt2 \epsilon_2 \mathrm{e}^{-i \phi/2} \sin x  &  -i \sqrt2 \epsilon_2\mathrm{e}^{i \phi/2} \cos x & 1
  \end{array} \right) +O(\epsilon^2_{1,2})\label{eq:CKMallorder}
   }
where $ x= \frac{\phi}{2} - \mathrm{sin}^{-1}\frac{\epsilon_1}{2}$. This CKM matrix is an interesting one since $V^{\rm CKM}_{31}$ is much smaller than
$\epsilon_2$ due to the small value of $\sin x$. A next order breaking term will generate a still smaller $V^{\rm CKM}_{13}$ and the form (\ref{eq:CKMallorder}) is phenomenologically a useful one \cite{Wolfenstein}. The phase redefinition of quarks with $\mathrm{diag}(\mathrm{exp}(i \phi/2), \mathrm{exp}(i \phi/2), 1 )$ or other symmetry breaking interaction could be used to obtain the form more close to the observed CKM matrix.

Similarly for the lepton sector, a $D_{12}$ violating term
\dis{ \epsilon (-y^l_1 H^l_0 + y^l_4 H^{\prime l}_0 + y^5_l H^{\prime l}_2 ) \bar{e}_L (\mu_R+ \tau_R ) + h.c.\label{eq:Dviolep}
}
gives the following the lepton mass matrix
  \dis{
  M^{(l)}
  =\left (\begin{array}{ccc}
   x_l & \frac{\epsilon}{\sqrt2}(-x_l+w_l+z_l) &  \frac{\epsilon}{\sqrt2}(-x_l+w_l+z_l) \\
  \frac{\epsilon}{\sqrt2}(-x_l+w_l+z_l) & w_l &  z_l \\
   \frac{\epsilon}{\sqrt2}(-x_l+w_l+z_l) & z_l & w_l
  \end{array} \right)
  }
which is diagonalized by
\dis{
  U_l\simeq \left (\begin{array}{ccc}  1 &0 & 0 \\
 0 & \frac{1}{\sqrt2} &  \frac{1}{\sqrt2} \\  0 & -\frac{1}{\sqrt2} & \frac{1}{\sqrt2} \end{array} \right)
  \left (\begin{array}{ccc} 1 &0 & \epsilon \\
 0 & 1 &  0 \\  - \epsilon & 0 & 1 \end{array} \right)
 \simeq\left (\begin{array}{ccc}  1 &0 & \epsilon \\
\frac{-\epsilon}{\sqrt2} & \frac{1}{\sqrt2} &  \frac{1}{\sqrt2} \\
  \frac{-\epsilon}{\sqrt2} & \frac{-1}{\sqrt2} & \frac{1}{\sqrt2} \end{array} \right) +O(\epsilon^2)
  }
In this case, the PMNS matrix is given by
\dis{
V_{\rm PMNS} = U_l^{\dagger}U_{\nu} =
  \left ( \begin{array}{ccc}
   \mathrm{cos} \frac{\pi}{6} + \frac{\epsilon}{\sqrt2} \sin \frac{\pi}{6}, & \mathrm{sin}\frac{\pi}{6} - \frac{\epsilon}{\sqrt2} \mathrm{cos} \frac{\pi}{6},  & - \frac{\epsilon}{\sqrt2}  \\
   - \frac{1}{\sqrt{2}} \mathrm{sin} \frac{\pi}{6}, & \frac{1}{\sqrt{2}}
   \mathrm{cos} \frac{\pi}{6}, & - \frac{1}{\sqrt{2}}  \\
    - \frac{1}{\sqrt{2}} \mathrm{sin} \frac{\pi}{6} + \epsilon \mathrm{cos} \frac{\pi}{6}, & \frac{1}{\sqrt{2}} \mathrm{cos}\frac{\pi}{6}
    + \epsilon  \mathrm{sin} \frac{\pi}{6}, & \frac{1}{\sqrt{2}}
  \end{array} \right) +O(\epsilon^2).
}
\end{widetext}
from which we notice that $\theta^{\rm PMNS}_{13}=- \frac{\epsilon}{\sqrt2}$
which is of order $\epsilon$. One way of constructing such a $ D_{12} $ violating term is to introduce very heavy particles coupled to the Higgs. For example, the Cabbibo angle shifting $ \epsilon_1 $ term implies the existence of $ \textbf{1}_{--} $ field from the relative minus sign between $ \bar{u}_L c_R $ and $ \bar{c}_L u_R $. For $ \epsilon_2 $ in the quark sector and $ \epsilon$ in the lepton sector, $ \textbf{2}_{1} $ fields with the VEVs proportional to $ x-z-w $ were introduced, $ y_1^u H_0^u - y_4^u H^{\prime u}_0 - y_5 H^{\prime u}_2 $ and $ y_1 H_0^l - y_4 H^{\prime l}_0 - y_5 H^{\prime l}_2 $ in (\ref{eq:Dvioqu}) and  (\ref{eq:Dviolep}), respectively. But it is unclear how these VEVs could be fine-tuned up to this order of $\epsilon^2$.

\section{Conclusion}\label{sec:Conclusion}
In this paper, we studied the discrete symmetry $D_{12}$ at the electroweak scale to fix the quark and lepton mixing angles.
A full symmetry we discussed beyond the SM gauge group is  $D_{12}\times $U(1)$_\Gamma\times Z_3\times Z_2$ where U(1)$_\Gamma$ is a PQ symmetry.
The Cabibbo angle is known to be small as schematically presented in Eq. (\ref{eq:mixgeneral}). The philosophy for obtaining this small angle is {\it \`a la} BHL where the phase angles are represented as multiples of $360^{\rm o}/(\rm integer)$. At the leading order, the Cabbibo angle $\theta_C$ is 15$^{\rm o}$ whence (integer) is chosen as 24. This is possible with a dodeca-symmetry, using the half angle formula of the cosine function. Of course, the entries of the PMNS matrix has phase angles which are multiples of 15$^{\rm o}$, leading to the Solar-neutrino angle $\theta_{\rm sol}=30^{\rm o}$ and  $\theta_{\mu\tau}=45^{\rm o}$. Thus, there results the relation $\theta_{\rm sol}+\theta_C\simeq 45^{\rm o}$. The resulting electroweak scale quark masses at the vacuum we chose has a non-vanishing Arg.Det.$M_q$ and there is a need to solve the strong CP problem by a PQ symmetry broken at the axion window \cite{KimRMP10} since the other possibility $m_u=0$ is ruled out by Manohar and Sachrajda in Ref. \cite{PData08}. Out of discrete vacua, a certain vacuum is chosen for this assignment to be consistent with the dodeca-symmetry. We also argued for a small breaking of the dodeca-symmetry to shift our leading term of $\theta_C=15^{\rm o}$ to the observed value of 13.14$^{\rm o}$  and to generate the small but nonzero values of $\theta^{\rm CKM}_{23}$ and $\theta^{\rm CKM}_{32}$, and the  smaller values of $\theta^{\rm CKM}_{13}$ and $\theta^{\rm CKM}_{31}$. This small next order breaking of the dodeca-symmetry also generates a small nonzero value of
 $\theta^{\rm PMNS}_{13}$.

The PMNS dodeca-form we presented here can be as attractive and phenomenologically successful as the much discussed tri-bimaximal form.

\acknowledgments{This work is supported in part by the National Research Foundation  (NRF) grant funded by the Korean Government (MEST) (No. 2005-0093841), and MS is supported in addition by Grant No. KRF-2008-313-C00162 and the FPRD of the BK21 program.}

\vskip 0.5cm
\centerline{\bf Appendix: A review of $ D_{2N} $ symmetry}
\vskip 0.2cm
The dihedral group $ D_{2N} $ represents the symmetry of a regular polygon of $2N$ sides. Its properties are:
\begin{enumerate}
\item  It is isomorphic to $ Z_{2N} \rtimes Z_2 $ (cyclic rotation + reflection).
\item  It is generated by two generators $a$ and $b$,
 \dis{
 & a : (x_1 , x_2 , \cdots , x_{2N} )  \to (x_{2N}, x_1, \cdots , x_{2N-1})
 \\
 & b : (x_1 , x_2 , \cdots , x_{2N} )  \to (x_1, x_{2N}, \cdots , x_2)
 }
 which satisfies
 \dis{
  a^{2N} = e, ~~ b^2 = e, ~~bab=a^{-1}.
 }
\item   Its irreducible representations are
\dis{
&{\rm Four~ singlets:} ~\boldsymbol{1} _{++}, \boldsymbol{1} _{--},
  \boldsymbol{1} _{+-},  \boldsymbol{1} _{-+}\\
& (N-1)-{\rm  doublets:}~ \boldsymbol{2} _k(k = 1,\cdots , N-1)
\label{rep:doubsing}
 }
For a (complex) $\boldsymbol{2} _k $ doublet basis, $a$ and $b$ are represented by
\dis{
\quad a= \left ( \begin{array}{cc}
   \mathrm{e}^{2 \pi i k /2N} & 0  \\
   0 & \mathrm{e}^{-2 \pi i k /2N}
  \end{array} \right),~~
   b= \left ( \begin{array}{cc}
   0 & 1  \\  1 & 0
  \end{array} \right)\label{eq:btrans}
}
For a (complex) $\boldsymbol{1} _{ij} $ singlet basis, $i$ is the eigenvalue of $b$ and $j$ is the eigenvalue of $ab$.

\item   Tensor products satisfy the following.

\begin{itemize}
 \item Singlet times  singlet multiplication,
\dis{
  \boldsymbol{1}_{s_1 s_2} \times \boldsymbol{1}_{s'_1 s'_2}
  = \boldsymbol{1}_{s''_1 s''_2}
}
where $ s''_1 = s_1 s_1 '  $ and $ s''_2 = s_2 s_2 '  $.
\item Singlet times doublet multiplication,
\dis{
  (w)(\boldsymbol{1}_{++}) \times
  \left ( \begin{array}{c} x_1   \\   x_2
  \end{array} \right) (\boldsymbol{2}_{k})
  =   \left ( \begin{array}{c}  w x_1   \\  w x_2
  \end{array} \right)(\boldsymbol{2}_{k}), ~
    (w)(\boldsymbol{1}_{--}) \times
  \left ( \begin{array}{c}   x_1   \\   x_2
  \end{array} \right)(\boldsymbol{2}_{k})
  =   \left ( \begin{array}{c}  w x_1   \\  -w x_2
  \end{array} \right) (\boldsymbol{2}_{k}),
}
\dis{  (w)(\boldsymbol{1}_{+-}) \times
  \left ( \begin{array}{c}   x_1   \\   x_2
  \end{array} \right)(\boldsymbol{2}_{k})
  =   \left ( \begin{array}{c}  w x_2   \\  w x_1
  \end{array} \right) (\boldsymbol{2}_{k}),   (w)(\boldsymbol{1}_{-+}) \times  \left ( \begin{array}{c}
   x_1   \\   x_2  \end{array} \right)(\boldsymbol{2}_{k})
  =   \left ( \begin{array}{c}  w x_2   \\  -w x_1
  \end{array} \right)(\boldsymbol{2}_{k}).
}
where the boldface symbols inside the brackets show the $D_{2N}$ representations.
\item Doublet times doublet multiplication,
\begin{widetext}

{(a)}  For $ k+k' \neq N $ and $ k-k' \neq 0 $,
     \begin{equation}
  \left ( \begin{array}{c}
   x_1   \\   x_2
  \end{array} \right) (\boldsymbol{2}_{k})  \times
    \left ( \begin{array}{c}   y_1   \\   y_2
  \end{array} \right) (\boldsymbol{2}_{k'})
 =   \left ( \begin{array}{c}   x_1 y_1   \\   x_2 y_2
  \end{array} \right) (\boldsymbol{2}_{k+k'})  +
   \left ( \begin{array}{c}   x_1 y_2   \\   x_2 y_1
  \end{array} \right) (\boldsymbol{2}_{k-k'}).\label{eq:sgsgtosing}
 \end{equation}
{(b)}    For $ k+k' = N $ and $ k-k' \neq 0 $ ,
     \begin{equation}
  \left ( \begin{array}{c}  x_1   \\   x_2
  \end{array} \right) (\boldsymbol{2}_{k}) \times
    \left ( \begin{array}{c}   y_1   \\   y_2
  \end{array} \right) (\boldsymbol{2}_{k'})
  =(x_1 y_1 + x_2 y_2)(\boldsymbol{1}_{+-})  +
(x_1 y_1 - x_2 y_2)(\boldsymbol{1}_{-+}) +
   \left ( \begin{array}{c}   x_1 y_2   \\   x_2 y_1
  \end{array} \right) (\boldsymbol{2}_{k-k'})\label{eq:sgdbtosing}
 \end{equation}
{(c)}   For $ k+k' \neq N $ and $ k-k' = 0 $ , (which will be frequently used)
     \begin{equation}
  \left ( \begin{array}{c}   x_1   \\   x_2
  \end{array} \right)(\boldsymbol{2}_{k})
  \times    \left ( \begin{array}{c}   y_1   \\  y_2
  \end{array} \right)(\boldsymbol{2}_{k'})
    =  (x_1 y_2 + x_2 y_1)(\boldsymbol{1}_{++})+
  (x_1 y_2 - x_2 y_1)(\boldsymbol{1}_{--}) +
   \left ( \begin{array}{c}
   x_1 y_1   \\   x_2 y_2
  \end{array} \right) (\boldsymbol{2}_{k+k'}).\label{eq:dbdbtosinga}
 \end{equation}
{(d)}  For $ k+k' = N $ and $ k-k' = 0 $ ,
\dis{
  \left ( \begin{array}{c}   x_1   \\  x_2
  \end{array} \right) (\boldsymbol{2}_{k})  \times
    \left ( \begin{array}{c}   y_1   \\   y_2
  \end{array} \right) (\boldsymbol{2}_{k'})
  =& (x_1 y_2 + x_2 y_1)(\boldsymbol{1}_{++}) +
  (x_1 y_2 - x_2 y_1)(\boldsymbol{1}_{--})\\
  &+(x_1 y_1 + x_2 y_2)(\boldsymbol{1}_{+-})
  +(x_1 y_1 - x_2 y_2)(\boldsymbol{1}_{-+}).\label{eq:dbdbtosingb}
}
\end{widetext}
\end{itemize}

\end{enumerate}

The symmetry breaking of $D_{2N}$ has been extensively discussed in Ref. \cite{Blum08}. The spontaneous symmetry is usually achieved by giving VEVs to Higgs scalar fields. For a  $D_{2N}$ doublet, the VEV is chosen as
\dis{
    \langle H(\boldsymbol{2}_{k}) \rangle \sim
  \left ( \begin{array}{c}
   \mathrm{e} ^{ \frac{-2 \pi i}{2N} k m}  \\   1
  \end{array} \right).\label{eq:HVEVapp}
}
Note that $ \langle H(\boldsymbol{2}_{k})\rangle $ is the eigenvector of $ ba^m $ with eigenvalue 1, and hence it is still invariant under the action of $ ba^m $. Therefore, by the VEV of Eq. (\ref{eq:HVEVapp}) $ D_{2N} $ is broken down to the smaller group generated by $ ba^m $. Since $ (ba^m)^2 =1$, the remaining group should have a subgroup $ Z_2 $ generated by $ba^m$. The symmetry breaking pattern for this vacuum choice is as follows:
\begin{itemize}
 \item When $ j $ divides $ 2N $ ($ m=0, 1, \cdots \frac{2N}{j}-1 $), $D_{2N}$ is broken down to
\dis{
 D_{2N} \stackrel{\textbf{2}_j}{\longrightarrow} D_j
 = \langle a^{2N/j}, ba^m \rangle .
}
Note that $ a^{2N/j} $ generates $ Z_j $ since $ (a^{2N/j})^j =1 $. Therefore, the group generated by $a^{2N/j}, ba^m $ is $ Z_j \rtimes Z_2 = D_j $.
\item When $ j $ does not divide $ 2N $ ($ m=0, 1, \cdots, 2N-1 $), $ D_{2N}$ is broken down to
\dis{
 D_{2N} \stackrel{\textbf{2}_j}{\longrightarrow}  Z_2 = \langle  ba^m \rangle
}
\item  A successive application of doublet VEVs lead to
(a) When $k$ divides $j$ with $m_j=m_k $,
\dis{
 D_{2N} \stackrel{\textbf{2}_j}{\longrightarrow} D_j \stackrel{\textbf{2}_k}{\longrightarrow} D_k.
}
 (b) When $k$ does not divide $j$ with $m_j=m_k $,
\dis{
  D_{2N} \stackrel{\textbf{2}_j}{\longrightarrow} D_j \stackrel{\textbf{2}_k}{\longrightarrow} Z_2.
}
\end{itemize}
Of course, one can choose an arbitrary value for the VEV, and Ref \cite{Blum08} lists all the possible symmetry breaking patterns and the resulting subgroups.

\end{document}